\documentclass[12pt]{article}
\usepackage{epsfig}
\usepackage{amsmath}
\usepackage{hhline}
\usepackage{amssymb}
\usepackage{times}
\usepackage{cite}

\newlength{\dinwidth}
\newlength{\dinmargin}
\renewcommand{\L}{{\cal L}}

\newcommand{\rpv}{\slash\hspace{-2.5mm}{R}_{p}}

\newcommand{\GeV}{\mathrm{GeV}}
\newcommand{\TeV}{\mathrm{TeV}}
\newcommand{\pb}{\mathrm{pb}}

\newcommand{\m}{\mathrm{m}}
\newcommand{\hdick}{\noalign{\hrule height1.4pt}}
\newcommand{\nn}{\nonumber}
\newcommand{\beq}{\begin{equation}}
\newcommand{\eeq}{\end{equation}}
\newcommand{\bea}{\begin{eqnarray}}
\newcommand{\eea}{\end{eqnarray}}
\newcommand{\eq}[1]{eq.~(\ref{#1})}

\begin{document}

\pagenumbering{roman}
\begin{titlepage}

\noindent
        
\begin{flushleft}
  {\tt DESY 03-052} \hfill {\tt ISSN 0418-9833} \\
  {\tt May 2003}
\end{flushleft}

\vspace*{2cm}

\begin{center}
  \begin{Large}{\bf\boldmath                   
      Search for new physics \\  in  
      $e^\pm q$ contact interactions at HERA}
    \\ \vspace*{2cm}
    {H1 Collaboration}
  \end{Large} 
\end{center}

\vspace*{2cm}

\begin{abstract} \noindent
Deep-inelastic $e^\pm p$ scattering at high squared
momentum transfer $Q^2$ up to $30000~\GeV^2$ 
is used to search for $eq$ contact interactions associated to scales 
far beyond the {\sc Hera} centre of mass energy.
The neutral current cross section measurements ${\rm d} \sigma / {\rm d} Q^2$,
corresponding to integrated luminosities of $16.4~\pb^{-1}$ of $e^-p$ data
and $100.8~\pb^{-1}$ of $e^+p$ data,
are well described by the Standard Model and 
are analysed to set constraints on new phenomena.
For conventional contact interactions lower limits are set on
compositeness scales $\Lambda$ ranging between $1.6 - 5.5~\TeV$.
Couplings and masses of leptoquarks and squarks in $R$-parity violating
supersymmetry are constrained to $M/\lambda > 0.3 - 1.4~\TeV$.
A search for low scale quantum gravity effects in models with large extra 
dimensions provides limits on the effective Planck scale of $M_S > 0.8~\TeV$.
A form factor analysis yields a bound on the radius of  light quarks of
$R_q < 1.0\cdot 10^{-18}$~m.
\end{abstract}

\vspace*{2.0cm}

\begin{center}
  {\it Submitted to Physics~Letters~B}
\end{center}

\vfill
\newpage
\begin{flushleft} \small 


C.~Adloff$^{33}$,              
V.~Andreev$^{24}$,             
B.~Andrieu$^{27}$,             
T.~Anthonis$^{4}$,             
A.~Astvatsatourov$^{35}$,      
A.~Babaev$^{23}$,              
J.~B\"ahr$^{35}$,              
P.~Baranov$^{24}$,             
E.~Barrelet$^{28}$,            
W.~Bartel$^{10}$
% ...

\bigskip{\it 
% ...
$ ^{4}$ Inter-University Institute for High Energies ULB-VUB, Brussels;
          Universiteit Antwerpen (UIA), Antwerpen; Belgium$^{ c}$ \\
% ...
$ ^{10}$ DESY, Hamburg, Germany \\
% ...
$ ^{23}$ Institute for Theoretical and Experimental Physics,
          Moscow, Russia$^{ l}$ \\
 $ ^{24}$ Lebedev Physical Institute, Moscow, Russia$^{ e}$ \\
% ...
$ ^{27}$ LPNHE, Ecole Polytechnique, IN2P3-CNRS, Palaiseau, France \\
 $ ^{28}$ LPNHE, Universit\'{e}s Paris VI and VII, IN2P3-CNRS,
          Paris, France \\
% ...
$ ^{33}$ Fachbereich Physik, Bergische Universit\"at Gesamthochschule
          Wuppertal, Wuppertal, Germany \\
% ...
$ ^{35}$ DESY, Zeuthen, Germany \\
% ...
$ ^c$ Supported by FNRS-FWO-Vlaanderen, IISN-IIKW and IWT \\
% ...
$ ^e$ Supported by the Deutsche Forschungsgemeinschaft \\
% ...
$ ^l$ Partially Supported by Russian Foundation
      for Basic Research, grant    no. 00-15-96584 \\
}

\end{flushleft}

\end{titlepage}

\pagenumbering{arabic}

\section{Introduction}

Deep inelastic neutral current scattering $e p \rightarrow e X$ at very high 
squared momentum \mbox{transfer $Q^2$} allows one to study the
structure of $e q$ 
interactions at short distances and to search for new phenomena beyond the 
Standard Model (SM). The concept of four-fermion contact interactions (CI) 
provides a convenient method to investigate the interference of any new 
particle field associated to large scales with the $\gamma$ and $Z$ fields of 
the Standard Model.
This paper considers conventional contact interactions, 
such as general models of compositeness and the exchange of heavy leptoquarks
and supersymmetric quarks,
as well as low scale quantum gravity effects, which may be mediated
via gravitons coupling to Standard Model particles
and propagating into large extra spatial dimensions.

The present analysis is a continuation of previous studies~\cite{h1ci} 
based on $e^+p$ data~\cite{h1xsec}.
New H1 cross section data on $e^-p$ scattering~\cite{h1xe-p} and
on $e^+p$ scattering data~\cite{h1xe+p} lead to a
substantially higher sensitivity to new physics phenomena.
Similar studies of $e q$ contact interactions have been performed by
other experiments at {\sc Hera}~\cite{zeusci}, {\sc Lep~\cite{lepci}} 
and {\sc Tevatron}~\cite{tevatronci}, providing results 
comparable to those presented here.

\section{Data and analysis method}

The data have been collected with the H1 detector at {\sc Hera} and 
correspond altogether to an integrated luminosity of 
$\L_{int} = 117.2~\pb^{-1}$. 
They consist of three data sets of different lepton charge and
centre of mass energy $\sqrt{s}$:
\begin{center} \vspace*{-2mm}
\begin{tabular}{ccccc}
          reaction & $\L_{int}~[\pb^{-1}]$ & $\sqrt{s}~[\GeV]$ 
        & run period & ref. \\[.2ex]  \hdick  \\[-2.3ex]                
          $e^+p \to e^+X$ & $35.6\pm0.53$ & 301
        & $1994 - 1997$ & \cite{h1xsec} \\[.2ex]
          $e^-p \to e^-X$ & $16.4\pm0.3$ & 319
        & $1998 - 1999$ & \cite{h1xe-p} \\[.2ex]
          $e^+p \to e^+X$ & $65.2\pm0.95$ & 319
        & $1999 - 2000$ & \cite{h1xe+p} \\
\end{tabular}
\end{center}
The cross section measurements ${\rm d}\sigma / {\rm d}Q^2$
extend over a large range of squared momentum 
transfers $200\,\GeV^2< Q^2 < 30000\,\GeV^2$ for inelasticity $y < 0.9$.
Details can be found in the quoted references.
In this paper
a contact interaction analysis is presented for each of the two recent
data sets and the combined data including the previous measurements.

The phenomenological models under study and the analysis method are 
described in more detail in ref.~\cite{h1ci}. 
The analysis investigates the measured cross sections
${\rm d}\sigma / {\rm d}Q^2$ and performs 
quantitative tests of the SM or CI models, applying a minimisation of the 
$\chi^2$ function
\begin{equation}
  \chi^2 = \sum_i \left( \frac {\hat{\sigma}_i^{\rm exp}-
      \hat{\sigma}_i^{\rm th}\, (1-\sum_k \Delta_{ik}(\varepsilon_k)) }
      {\Delta\hat{\sigma}_i} \right)^2 + \sum_k \varepsilon_k^2 \ .
  \label{chi2}
\end{equation}
Here $\hat{\sigma}_i^{\rm exp}$ and $\hat{\sigma}_i^{\rm th}$ are the
experimental and theoretical cross sections for the measurement point $i$
and $\Delta\hat{\sigma}_i$ is the corresponding error including statistical
and uncorrelated systematic errors added in quadrature.
The functions $\Delta_{ik}(\varepsilon_k)$ describe correlated systematic
errors for point $i$ associated to a source $k$. 
They depend on the fit parameters $\varepsilon_k$, which are effectively
pulls caused by systematics.
In general the influence of the correlated systematic errors is small.
The following dominant sources of correlations are included:
an overall normalisation uncertainty of $1.5 - 1.8\,\%$ depending on the
run period,
experimental uncertainties on the scattered lepton energy and angle
and an uncertainty on the strong coupling $\alpha_s(M_Z) = 0.118 \pm 0.003$ 
entering the SM prediction.
In a combined $\chi^2$ analysis using \eq{chi2} all data sets are treated 
as independent samples with individual normalisation and measurement 
uncertainties.
The influence on the fit results of correlations between data sets is
negligible. 

\begin{figure}[p] 
  \begin{center} \vspace*{-.5cm}
    \mbox{
      \epsfig{file=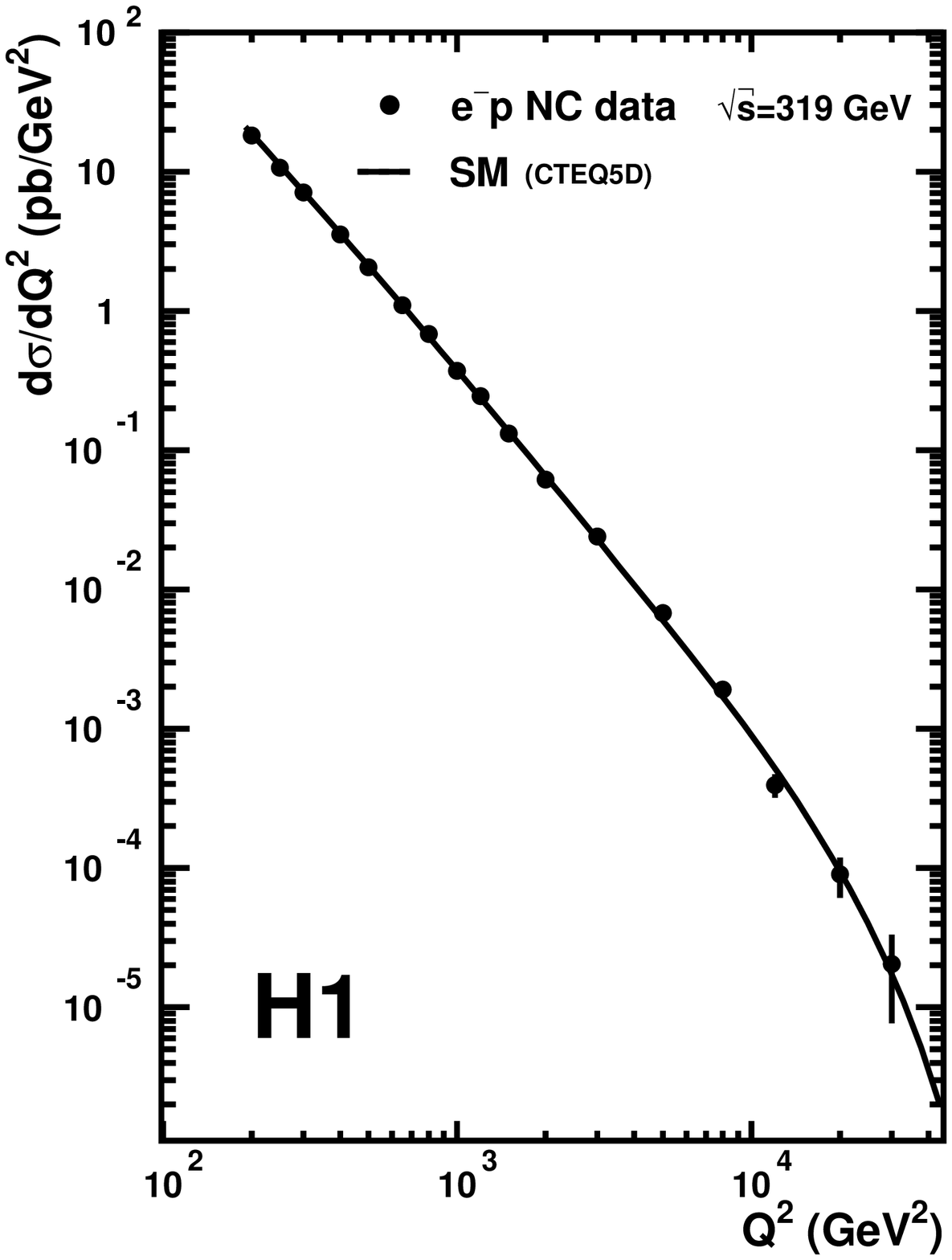,width=.49\textwidth,
        bbllx=0,bblly=0,bbury=570,bburx=420,clip}  \hfill
      \epsfig{file=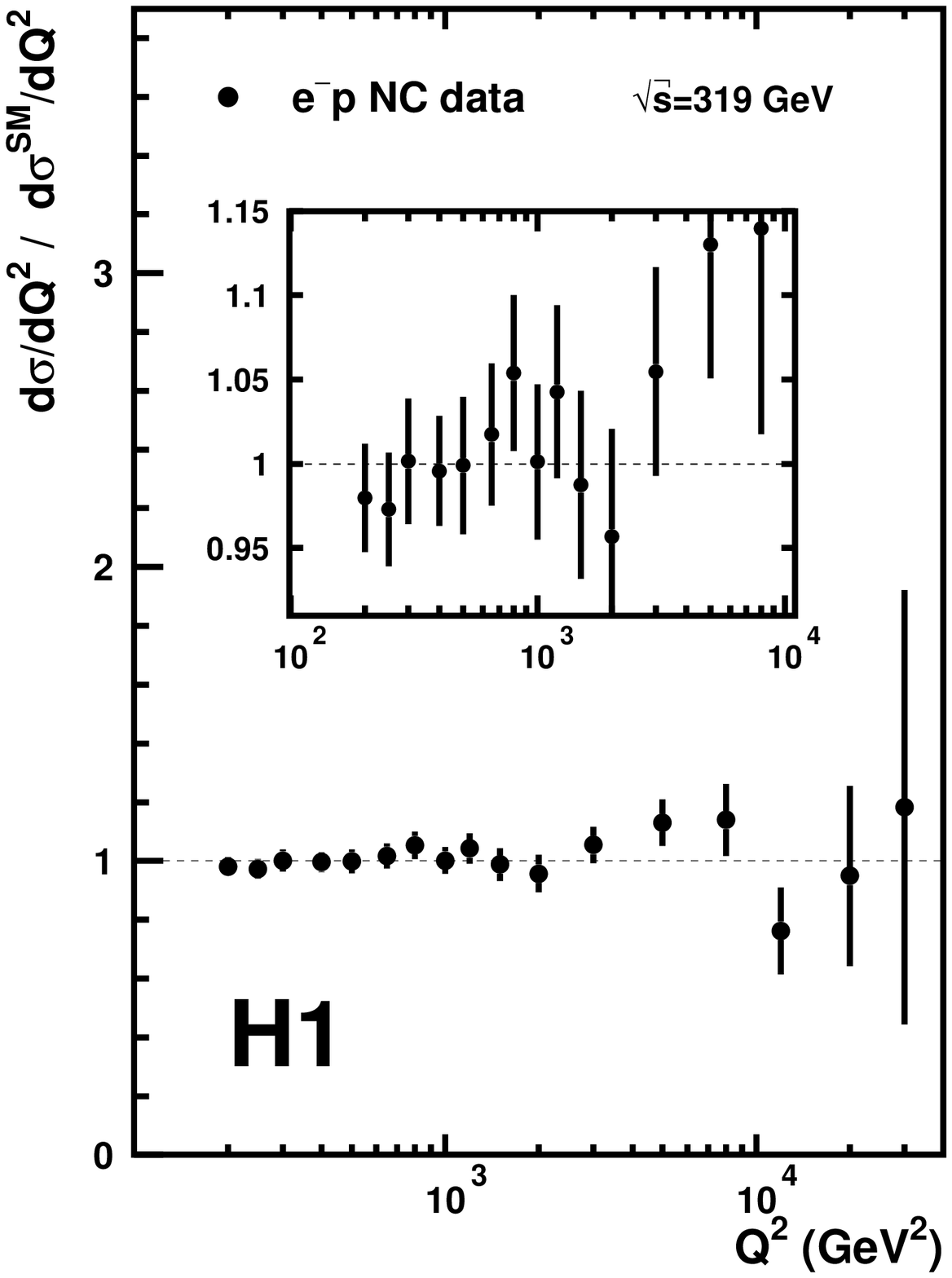,width=.49\textwidth,
        bbllx=0,bblly=0,bbury=570,bburx=420,clip}
      }\\[1em]
    \mbox{
      \epsfig{file=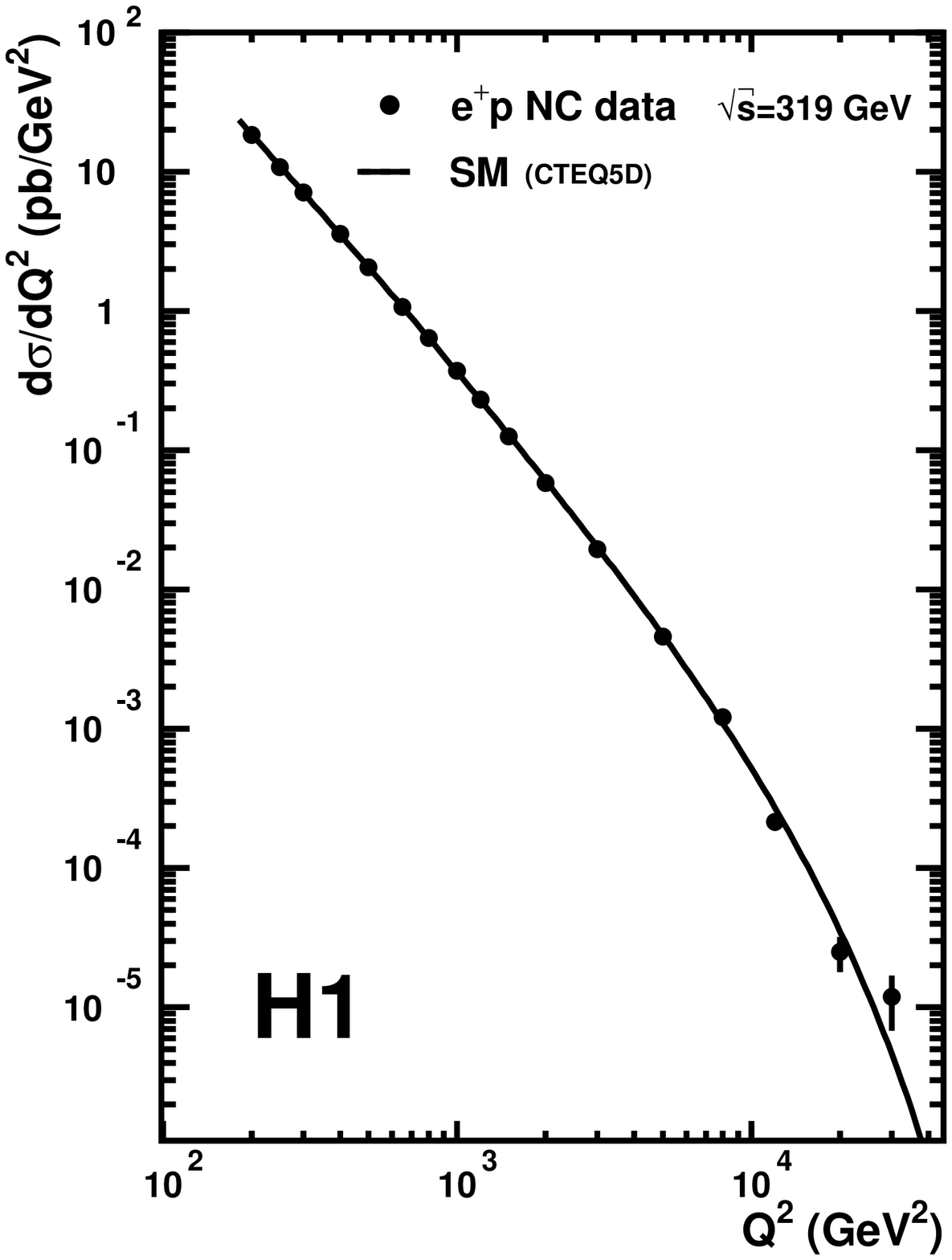,width=.49\textwidth,
        bbllx=0,bblly=0,bbury=570,bburx=420,clip}  \hfill
      \epsfig{file=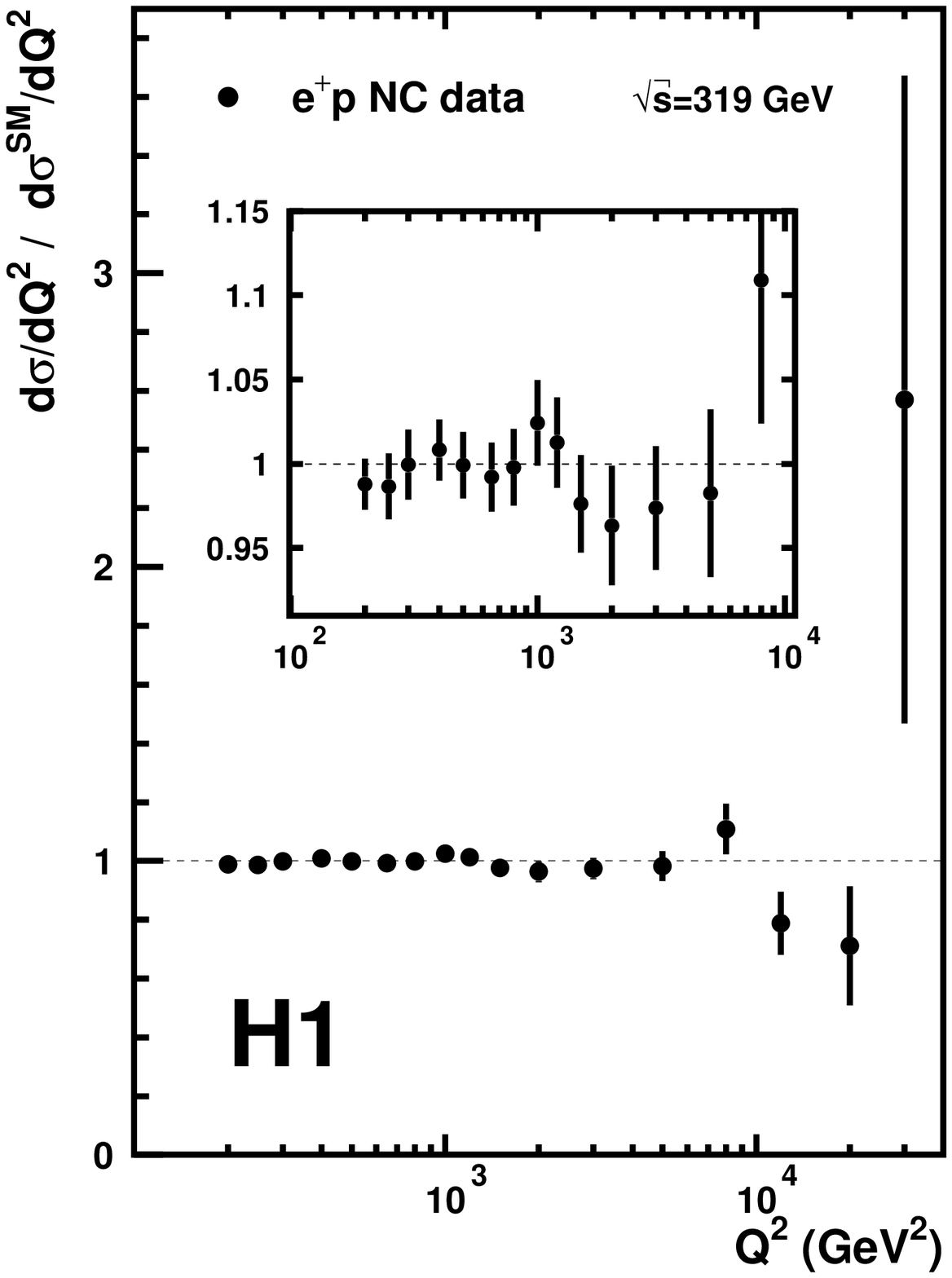,width=.49\textwidth,
        bbllx=0,bblly=0,bbury=570,bburx=420,clip}
      }
  \end{center} \vspace*{-.3cm}
  \caption{Cross sections
    ${\rm d}\sigma / {\rm d}Q^2$ at $\sqrt{s} = 319~\GeV$
    for $e^-p \rightarrow e^-X$ scattering (top) and 
    $e^+p \rightarrow e^+X$ scattering (bottom).
    H1 data are compared with Standard Model expectations 
    using the CTEQ5D parton distributions.
    The errors include only statistics and uncorrelated experimental systematics. 
    Normalisation uncertainties are 1.8\% ($e^-p$ data) 
    and 1.5\% ($e^+p$ data). }
  \label{cismxsec}
\end{figure}

The cross sections ${\rm d}\sigma (e^-p \to e^-X)/{\rm d}Q^2$~\cite{h1xe-p}
and ${\rm d}\sigma (e^+p \to e^+X)/{\rm d}Q^2$~\cite{h1xe+p} from the new data
are shown in figure~\ref{cismxsec}.\footnote{
  Corresponding figures for the $e^+p$ cross sections at 
  $\sqrt{s}=301~\GeV$~\cite{h1xsec} are shown in \cite{h1ci}. }
They are very well described 
by the Standard Model expectations 
over the full $Q^2$ range, over which
they vary by six orders of magnitude.
Cross section calculations in the Standard Model 
are performed in the DIS scheme in next-to-leading order (NLO) QCD
using the corresponding CTEQ5D parton parametrisation~\cite{cteq}.
These parton distributions were obtained
including the small amount of $2.7~\pb^{-1}$ 
of 1994 H1 data with $Q^2 < 5000~\GeV^2$ only  and are thus essentially
uncorrelated with the present data.

Fits to the SM prediction yield $\chi^2 = 10.4$ for $17$ degrees of
freedom (dof) with a fitted normalisation of $0.998$ for the $e^-p$ data and
$\chi^2/{\rm dof} = 13.9/17$ with a normalisation of $1.006$
for the $e^+p$ data, using statistical errors from the experiment
and including all systematics as in  \eq{chi2}.
Applying other parton density functions like CTEQ6~\cite{cteq6},
MRST~\cite{mrst} or GRV~\cite{grv}
leads to slight changes of the normalisation within errors, but
yields equally good agreement with the data.

The cross section measurements do not show significant deviations from the 
Standard Model and will be used to set limits on couplings from processes 
mediated through contact interactions.
Since the concept of contact interactions is an effective theory,
theoretical expectations cannot be formulated consistently in NLO. 
A sensible approach is to reweight at each $Q^2$ value
the SM NLO cross section by
$\hat{\sigma}_i^{LO}({\rm SM + CI})/\hat{\sigma}_i^{LO}({\rm SM})$,
the ratio of leading order (LO) cross sections with and without inclusion
of contact interactions. 
The $\chi^2$ function of \eq{chi2} can be applied 
to evaluate the sensitivity of the data to a certain CI scenario and
to determine its parameters.
The corresponding limits at 95\% confidence level (CL) are derived by 
using a frequentist approach as described in the Appendix.
Systematic uncertainties are included in this procedure.
This determination of limits is different from the method used in our previous 
publication~\cite{h1ci}.

\section{Contact interaction formalism}

The most general chiral invariant Lagrangian for neutral current vector-like  
four-fermion
contact interactions can be written in the form~\cite{elpr,haberl}
\begin{eqnarray}
  {\cal L}_V  &=& \sum_{q } 
  \sum_{a,\,b\, =\, L,\,R}
  \eta^q_{ab}\, (\bar{e}_a\gamma_\mu e_a)(\bar{q}_b\gamma^\mu q_b) \ .
 \label{lcontact}
\end{eqnarray}
For lepton $e$ and each {\em up}-type and {\em down}-type quark flavour $q$ 
with the corresponding currents $e_a$ and $q_b$
there are four coupling coefficients 
$\eta^q_{ab} = \epsilon_{ab}\, (g/\Lambda^q_{ab})^2$,
where $a$ and $b$ indicate the $L$ (left-handed)
and $R$ (right-handed) fermion helicities,
$g$ is the overall coupling strength, $\Lambda^q_{ab}$ is a scale parameter
and $\epsilon_{ab}$
determines the interference sign with respect to the Standard Model currents.
Any particular model, such as compositeness or the exchange of leptoquarks
or supersymmetric quarks,
can be constructed by an appropriate choice of the couplings $\eta^q_{ab}$.
The phenomenological models of interest and their analytical
treatment are discussed in more detail in~\cite{h1ci}.

\section{Compositeness scales}

In general models allowing for
fermion compositeness or substructure it is convenient
to choose a coupling strength of $g^2 = 4\,\pi$ and to assume a universal 
scale $\Lambda$ for all quarks.
The contact interaction coefficients then read
\begin{equation}
  \eta^q_{ab} = \epsilon_{ab}\,\frac{4\,\pi}{\Lambda^2} \ . \nn
\end{equation}
Various scenarios of chiral structures, e.g. pure L(eft) and R(ight)
or V(ector) and A(xial vector) couplings,
are defined by setting particular chiral contributions to values of
$\epsilon_{ab} = \pm 1$ and setting all other combinations to zero. 

The sensitivity to CI models is tested by determining the quantities
$\epsilon/\Lambda^2$ in a $\chi^2$ fit using the experimental statistical 
and systematic errors and leaving the sign of interference free.
The results from a combined fit to all data sets are shown in 
figure~\ref{cieta}.
The data tend to prefer negative interference,
but are consistent with the Standard Model within two standard 
deviations in all scenarios

\begin{figure}[htb]
  \begin{center}
     \epsfig{file=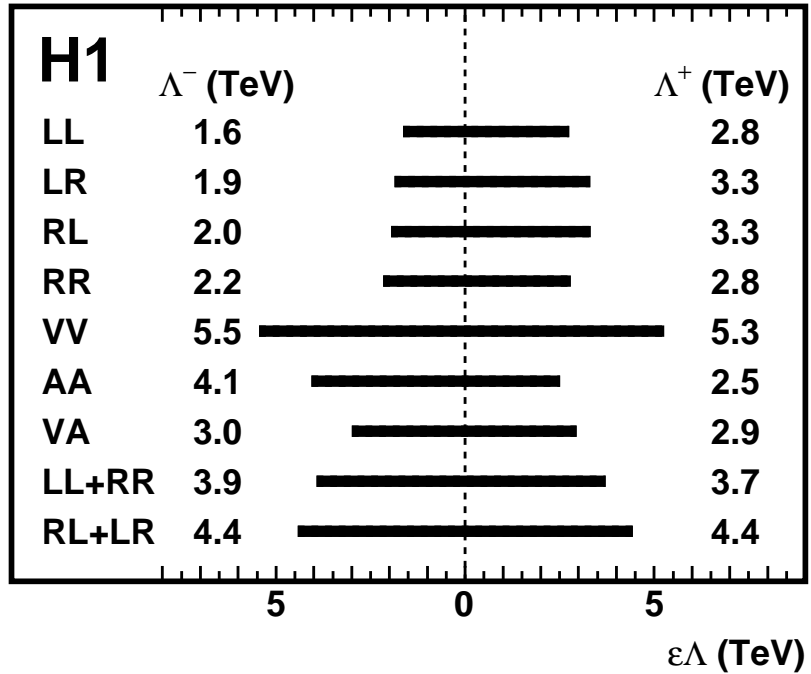,%
      bbllx=0,bblly=0,bbury=300,bburx=370,clip,%
      width=.7\textwidth}
  \end{center} \vspace{-5mm}
  \caption{Exclusion regions and
    lower limits (95\% CL) on compositeness scale parameters
    $\Lambda^\pm$ for various chiral models
    obtained from the combined $e^+ p$ and $e^- p$ data
    applying a frequentist method.}
  \label{cilambda}
\end{figure}

\begin{figure}[htb]
  \begin{center} 
      \epsfig{file=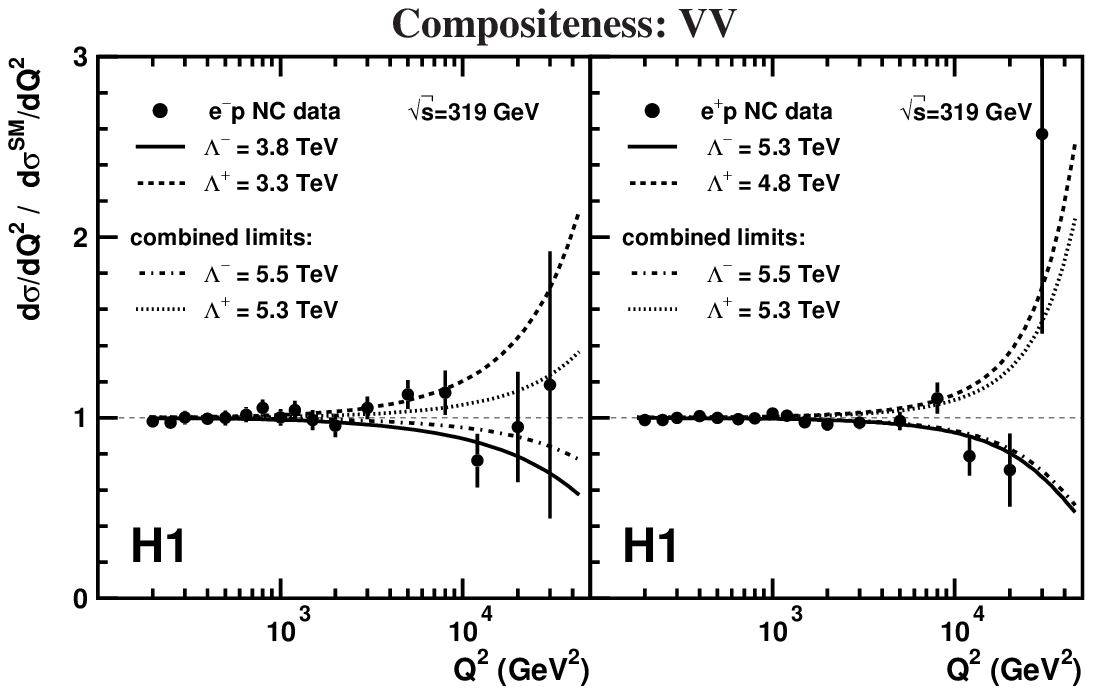,%
        bbllx=90,bblly=485,bbury=695,bburx=425,clip,%
        width=.7\textwidth} \\[-1ex]
      \epsfig{file=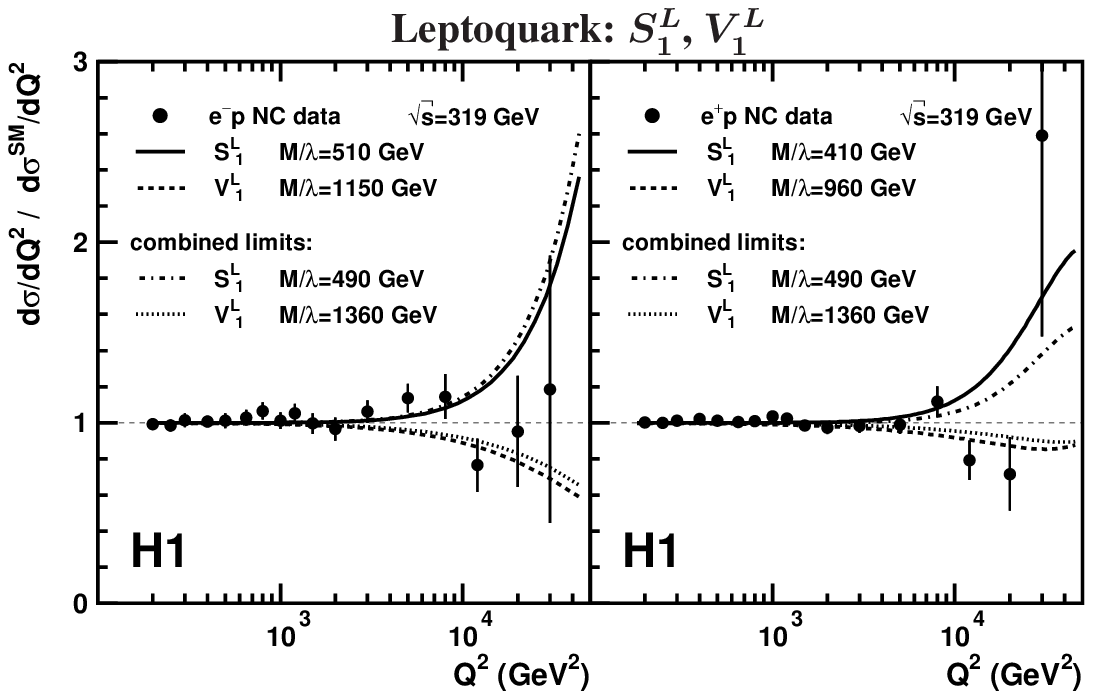,%
        bbllx=90,bblly=485,bbury=695,bburx=425,clip,%
        width=.7\textwidth} \\[-1ex]
      \epsfig{file=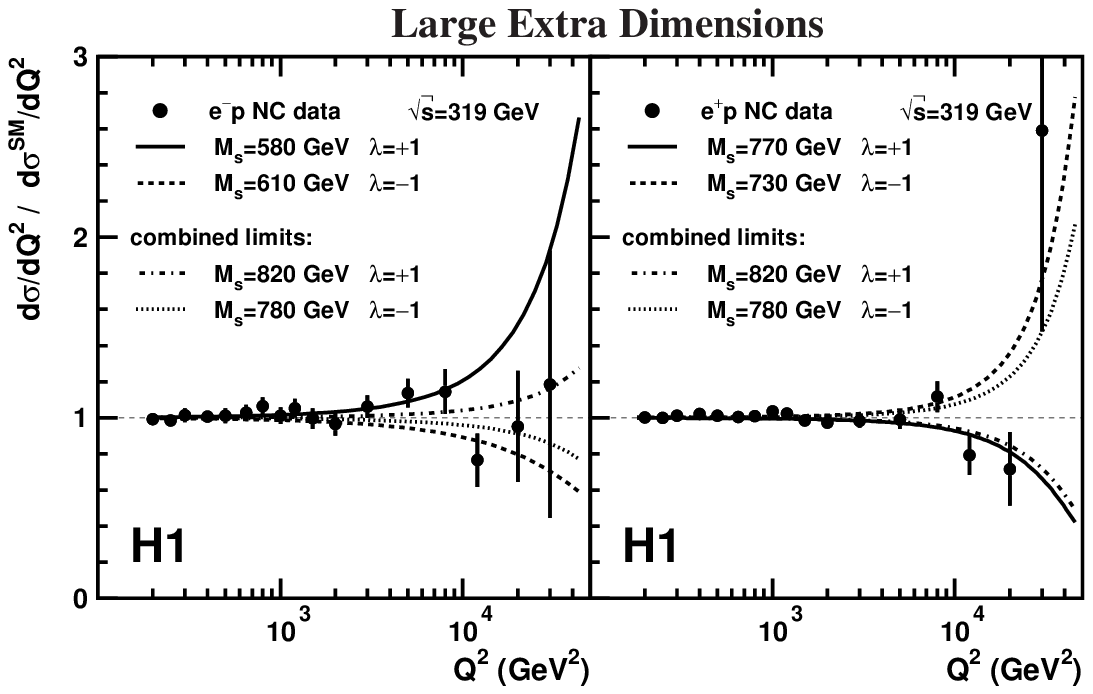,%
        bbllx=90,bblly=485,bbury=695,bburx=425,clip,%
        width=.7\textwidth}
  \end{center} \vspace*{-8mm}
  \caption{NC cross section ${\rm d}\sigma/{\rm d}Q^2$ at $\sqrt{s} = 319~\GeV$
    normalised to the Standard Model expectation.
    H1 $e^-p$ and  $e^+p$ scattering data
    are compared with curves corresponding to 95\%~CL exclusion limits 
    obtained from each data set and from the combined data for
    $VV$ compositeness scales $\Lambda^+$ and $\Lambda^-$ (top),
    couplings $M/\lambda$ of leptoquarks $S^L_{1}$ and $V^L_{1}$ (center),
    gravitational scales $M_S$ assuming positive ($\lambda = +1$) 
    and negative ($\lambda = -1$) couplings (bottom).
    The errors represent only statistics and uncorrelated experimental 
    systematics.}
  \label{civvxsec}
  \label{cilqxsec}
  \label{ledeffect}
\end{figure}

Lower limits on compositeness scale parameters $\Lambda^\pm$,
associated to positive or negative interference,
are derived from a frequentist approach and
are summarised in table~\ref{etafits}.
Despite the lower integrated luminosity, 
the $e^-p$ data exhibit for some models
comparable ($LL$ and $RR$ couplings) or even higher ($AA^+$ coupling)
sensitivity than the $e^+p$ data.
The two lepton charges complement each other and 
a combined analysis of all $e^\pm p$ data sets yields substantial
improvements in most scenarios
compared with the previous publication~\cite{h1ci}.
The results are presented in figure~\ref{cilambda}.
The lower limits on compositeness scale parameters vary between
$1.6~\TeV$ and $5.5~\TeV$ depending on the chiral structure.
Choosing different parton distributions~\cite{cteq6,mrst,grv}
changes the quoted limits typically by a few per~cent, at most by $15\,\%$.
The most restrictive bounds are observed for the $VV$ model, where all
chiral components contribute with the same sign.
As an illustration, examples of fits to the $e^-p$ and $e^+p$ cross sections 
are shown in figure~\ref{civvxsec}.

\section{Leptoquarks}

Leptoquarks couple to lepton--quark pairs and appear in extensions of the
Standard Model which try to connect the lepton and quark sectors. 
They are colour triplet scalar or vector bosons, carrying
lepton ($L$) and baryon ($B$) number and
a fermion number $F = L + 3\,B$.
Since $F  = 2$ for $e^- q$ and $F=0$ for $e^+ q$ states,
one expects different sensitivities to particular leptoquark types
from electron and positron scattering.
For high enough mass scales
the leptoquark mass $M_{LQ}$ and its coupling $\lambda$ are related to the
contact interaction coefficients via 
\begin{equation}
  \eta^{q}_{ab} = \epsilon^{q}_{ab}\, \frac{\lambda^2}{M_{LQ}^2} \ . \nn
\end{equation}
Within the model of~\cite{brw} the production and decay modes are fixed
and the relative contributions $\epsilon^{q}_{ab}$ have been 
calculated~\cite{kalinowski}.
The notation, quantum numbers and couplings of the various leptoquarks
are given in table~\ref{lqfits}.

The analyses of the
cross section measurements do not show an indication of a leptoquark signal.
The results of fits for each type of leptoquark
are interpreted in terms of limits on the ratio $M_{LQ}/\lambda$  
and are summarised in table~\ref{lqfits}. 
In some cases, e.g.~$S^L_1$ and $V^L_1$,
the $e^-p$ data give more restrictive bounds than $e^+p$ scattering despite
the lower integrated luminosity.
This sensitivity is illustrated in figure~\ref{cilqxsec}, which shows 
possible contributions of the leptoquarks $S^L_1$ and $V^L_1$ 
to the $e^-p$ and $e^+p$ cross sections.
The two leptoquarks differ by their spin and couple with the same chiral 
structure but different strength and sign to $u$ and $d$ quarks.
This emphasises the complementary role and importance of both electron
and positron beams.
The combined analyses of all $e^\pm p$ data 
further constrain the search for leptoquarks,
reaching exclusion values of up to $M_{LQ}/\lambda = 1.4~\TeV$.
Note that upper bounds on the coupling strength $\lambda$
can only be set for leptoquark masses exceeding the accessible centre of mass
energy of {\sc Hera}.

\section{Squarks in {\boldmath $R_p$} violating supersymmetry}

In the most general formulation of supersymmetry there exist operators
which couple a lepton-quark pair to a squark, the scalar superpartner of 
a quark.
Such couplings violate $R$-parity (via lepton number violation), 
defined as $R_p=(-1)^{3B+L+2S}$ with S being the spin.
Thus $R_p=+1$ for SM particles and $R_p=-1$ for superpartners. 
This interaction allows single squarks to be produced or exchanged 
in deep inelastic scattering~\cite{rpv} via
\begin{eqnarray}
  e^+ d_R & \to & \tilde{u}_L,\ \tilde{c}_L,\ \tilde{t}_L 
           \quad\quad \  {\rm coupling} \ \lambda'_{1j1} \ ,
           \label{l1j1} \\
  e^+ \bar{u}_L & \to & \bar{\tilde{d}}_R, \ \bar{\tilde{s}}_R, \ 
                        \bar{\tilde{b}}_R
           \quad\quad {\rm coupling} \ \lambda'_{11k} \ .
           \label{l11k} 
\end{eqnarray}
The subscripts $ijk$ of the coupling $\lambda'_{ijk}$ describe 
the generation indices of the left-handed leptons, the left-handed quarks and 
the right-handed {\em down}-type quarks of the superfields, respectively.

The $e^+q$ coupling of the left-handed {\em up}-type squark of
reaction~(\ref{l1j1}) is the same as that of the scalar leptoquark
$\tilde{S}^L_{1/2}$, 
and the coupling of the right-handed {\em down}-type squark of
reaction~(\ref{l11k}) is the same as that of the scalar leptoquark
$S^L_0$.
Therefore the formalism and results of the leptoquark analysis can be directly
applied.
Limits on the ratio $M_{\tilde q}/\lambda'$ for $R_p$ violating squarks,
assuming branching ratios ${\cal B}_{\tilde q \to e q} = 1$,
are given in table~\ref{sqfits}.
Note that the squark generations cannot be distinguished in this analysis.

\section{Large extra dimensions}

It has been suggested that the gravitational scale $M_S$ in $4+n$ dimensional
string theory may be as low as the electroweak scale of order TeV~\cite{add}.
The relation to the Planck scale $M_P\sim 10^{19}~\GeV$ and the size $R$ of the
$n$ compactified extra dimensions is given by $M_P^2 \sim R^n\,M_S^{2+n}$.
In some models with large extra dimensions the SM particles reside on a
four-dimensional brane, while the spin 2 graviton propagates 
into the extra spatial dimensions and appears in the four-dimensional
world as a tower of  
massive Kaluza-Klein states with a level spacing $\Delta m =1/R$. 
The gravitons couple to the SM particles via the energy-momentum tensor with
a tiny strength given by the inverse Planck scale.
However, the summation over the enormous number of Kaluza-Klein states
up to the ultraviolet cut-off scale, taken as $M_S$,
leads to an effective contact-type interaction~\cite{giudice} with a coupling
\begin{equation}
  \eta_G = \frac{\lambda}{M^4_S} \ . \nn
\end{equation}
The coefficient $\lambda$ depends on details of the theory 
and is expected to be of order unity.
However, by convention, one also allows for a negative coupling and
thus sets $\lambda = \pm 1$.
The scale dependence of gravitational effects is quite
different from that of conventional contact interactions
discussed in the previous sections.

The cross section formulae for 
virtual graviton exchange in DIS have been calculated
in~\cite{h1ci} using the conventions and formalism of \cite{giudice}.
The interference of the graviton with the photon and $Z$ fields
has opposite sign for electron and positron scattering,
as illustrated in figure~\ref{ledeffect}.
Lower limits on the scale parameter $M_S$, derived from fits to the
${\rm d}\sigma / {\rm d}Q^2$ distributions including gravitational 
contributions, are summarised in table~\ref{ledfits}.
There is similar sensitivity to the effective gravitational scale 
for positive and negative interference,
resulting in lower limits of $M_S > 0.82~\TeV$ for $\lambda =+1$ and
$M_S > 0.78~\TeV$  for $\lambda =-1$.

In other scenarios the gauge bosons $\gamma$ and $Z$ are also assumed to
propagate into extra dimensions. 
This leads to analogous Kaluza-Klein states which couple to matter with 
electroweak strength and interfere with the ordinary gauge boson fields.
In a specific model of $4+1=5$ dimensions~\cite{cheung} with
compactification radius $R=1/M_C$,
the sums over the exchange of Kaluza-Klein gauge bosons essentially modify 
the photon propagator $1/Q^2 \to 1/Q^2 + \pi^2/(3\,M^2_C)$ and
the $Z$ propagator $1/(Q^2+M^2_Z) \to 1/(Q^2+M^2_Z) +\pi^2/(3\,M^2_C)$.
A fit to the data yields $M_C > 1.0~\TeV$ as a lower limit at 95\%~CL for the 
compactification scale.

\section{Form factors}

A fermion substructure can also be formulated by assigning a finite
size to the electroweak charge distributions. 
It is convenient to introduce electron and quark
form factors $f(Q^2)$ which reduce the Standard Model cross section 
at high momentum transfer~\cite{koepp}:
\begin{eqnarray}
  f(Q^2) & = & 1 - \frac{\langle r^2 \rangle}{6}\,Q^2 \ ,  \\
  \frac{{\rm d}\sigma}{{\rm d}Q^2} & = &
  \frac{{\rm d}\sigma^{\rm SM}}{{\rm d}Q^2}\,f^2_e(Q^2)\,f^2_q(Q^2) \ .
\end{eqnarray}
Fits to the data yield upper limits on
the particle size $R = \sqrt{\langle r^2 \rangle}$, taken as the root
of the mean squared radius of the electroweak charge distribution.
Assuming a point-like electron, i.e. setting $f_e \equiv 1$, the radius of the
light $u$ and $d$ quarks can be constrained to $R_q < 1.0\cdot 10^{-18}~\m$
at 95\%~CL. If both electrons and quarks are assumed to have common
form factors one obtains limits on fermion sizes of 
$R_{e,\,q} < 0.7\cdot 10^{-18}~\m$.
The results of the analysis are given in table~\ref{rqfits}.

\section{Summary}

Neutral current deep inelastic $e^-p$ and $e^+p$ scattering
cross section measurements
are analysed to search for new phenomena mediated through 
$(\bar{e} e)\,(\bar{q} q)$ contact interactions.
The data are well described by the Standard Model expectations.
The use of electrons and positrons
provides complementary information and a combined analysis 
based on an integrated luminosity of $117.2~\pb^{-1}$
yields improved limits on scales of new physics.
The present analysis supersedes previous results~\cite{h1ci}.

Lower limits at 95\%~CL on $e q$ compositeness scale parameters $\Lambda^\pm$
are derived within a model independent analysis.
They range between $1.6~\TeV$ and $5.5~\TeV$ depending on the chiral structure.
A study of virtual leptoquark exchange yields lower limits on 
the ratio $M_{LQ}/\lambda$ between $0.3~\TeV$ and $1.4~\TeV$.
Squarks in $R$-parity violating supersymmetry with masses satisfying
$M_{\tilde u}/\lambda'_{1j1} < 0.43~\TeV$ and
$M_{\tilde d}/\lambda'_{11k} < 0.71~\TeV$ can be excluded.
Possible effects of low scale quantum gravity with gravitons 
propagating into extra spatial dimensions are searched for.
Lower limits on the effective Planck scale $M_S$ of $0.78 - 0.82~\TeV$ 
are found.
Allowing for Kaluza-Klein states of the SM gauge bosons results in a lower
limit on the extra dimension compactification scale $M_C > 1.0~\TeV$.
A form factor approach yields an upper limit on the size
of light $u$ and $d$ quarks of $R_q< 1.0\cdot 10^{-18}~\m$ 
assuming point-like electrons.

\paragraph{ Acknowledgements}
We are grateful to the {\sc Hera} machine group whose outstanding 
efforts have made this experiment possible. We thank 
the engineers and technicians for their work in constructing and now 
maintaining the H1 detector, our funding agencies for financial support, the 
{\sc Desy} technical staff for continual assistance, and the 
{\sc Desy} directorate for support and 
the hospitality extended to the non--{\sc Desy} members of the collaboration.

\appendix
\section*{Appendix \quad Setting limits}

For the present analysis several methods have been studied to calculate
limits and confidence levels and the final results quoted are based on the 
frequentist approach (see for example~\cite{pdg}).
The $\chi^2$ function of \eq{chi2} is used as the quality measure of agreement
between data and contact interaction models. 
It allows an easy implementation of systematic uncertainties.
The other methods investigated differ in the definitions of the statistical 
error entering the total error $\Delta\hat\sigma_i$ in the $\chi^2$ expression.

\begin{figure}[htb]
  \begin{center}\vspace*{-10mm}\hspace*{-20mm}
    \epsfig{file=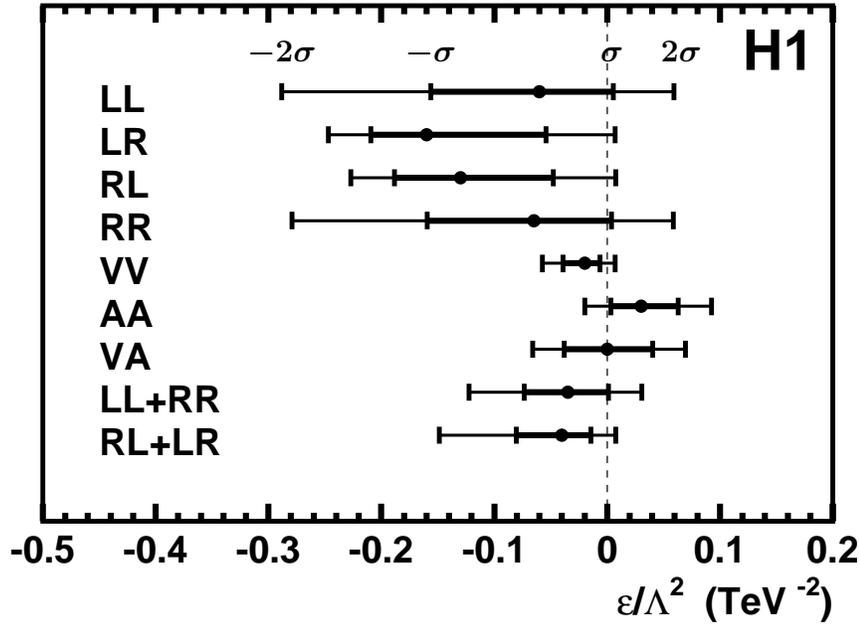,width=.8\textwidth}
    \setlength{\unitlength}{1mm}
    \begin{picture}(100,0) \boldmath
      \put(22,80){{$-2\sigma$}}  
      \put(43,80){{$-\sigma$}}  
      \put(69,80){{$\sigma$}}  
      \put(77,80){{$2\sigma$}}  \unboldmath
    \end{picture}
  \end{center}\vspace{-10mm}
  \caption{Fit results on the parameters $\epsilon/\Lambda^2$ 
    of compositeness models using the combined 
    $e^+ p$ and $e^- p$ data.
    The inner and outer error bars represent one and two standard deviations,
    respectively, 
    obtained by using experimental statistical and systematic errors.}
  \label{cieta}
\end{figure}

\begin{figure}
  \begin{center}
    \includegraphics[width=0.32\textwidth,%
    bbllx=0,bblly=0,bbury=278,bburx=278,clip]
    {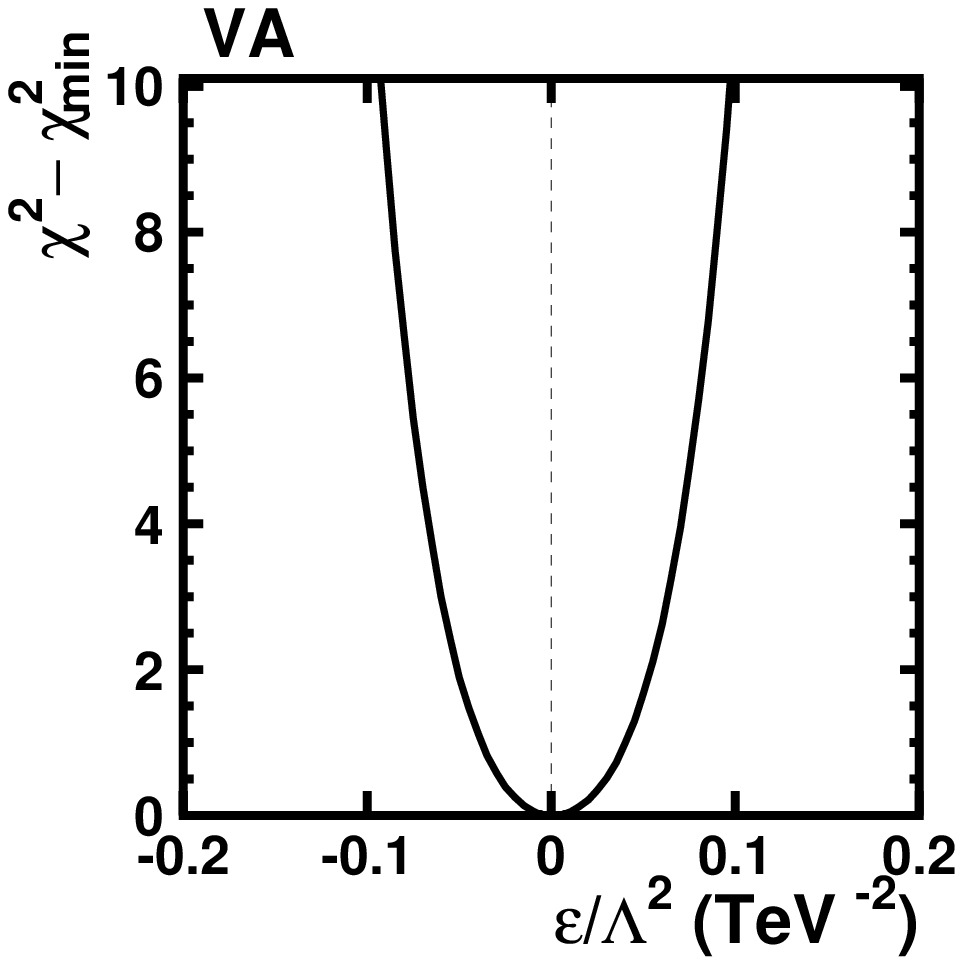}
    \includegraphics[width=0.32\textwidth,%
    bbllx=0,bblly=0,bbury=278,bburx=278,clip]
    {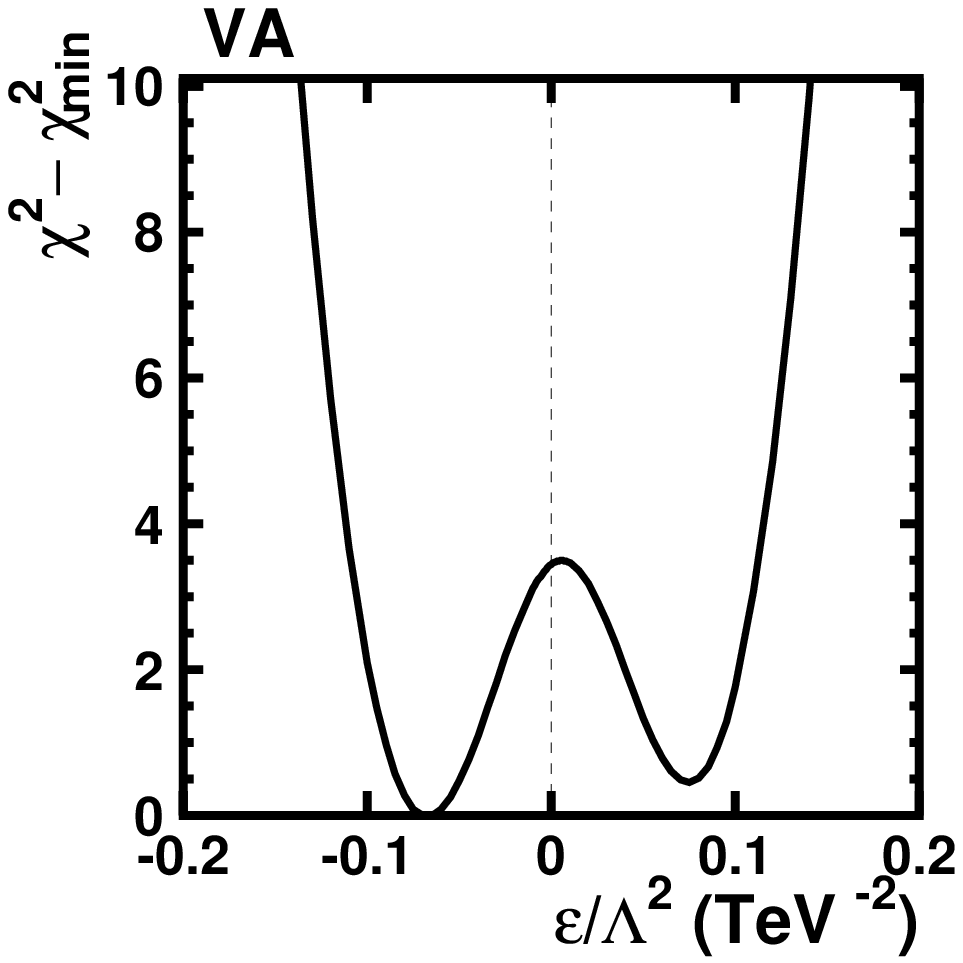}
    \includegraphics[width=0.32\textwidth,%
    bbllx=0,bblly=0,bbury=278,bburx=278,clip]
    {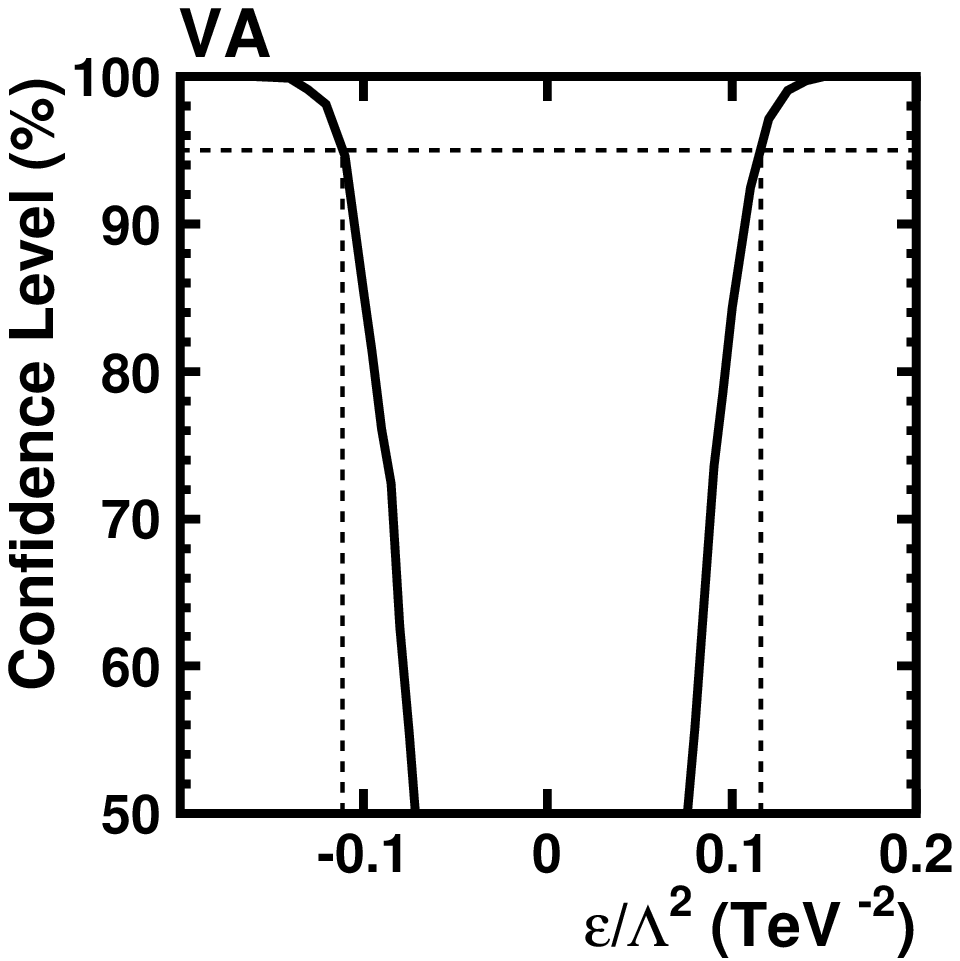}
    \\[5ex]
    \includegraphics[width=0.32\textwidth,%
    bbllx=0,bblly=0,bbury=278,bburx=278,clip]
    {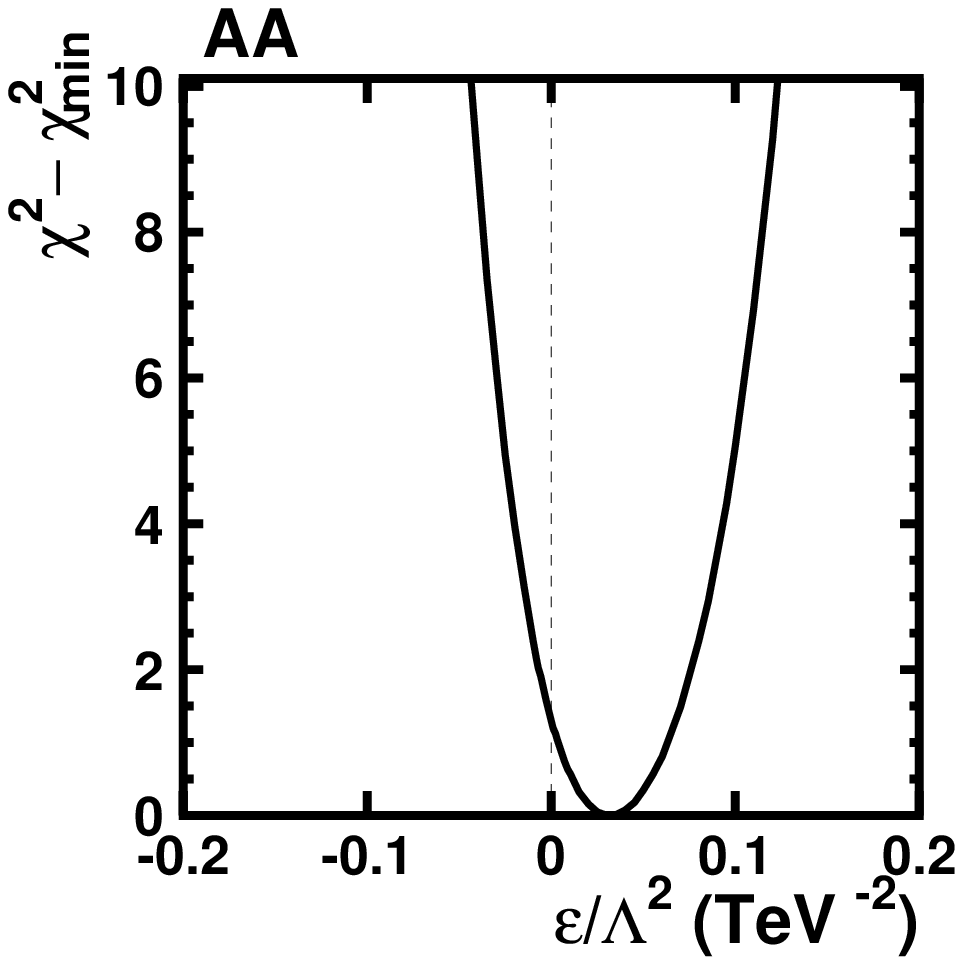}
    \includegraphics[width=0.32\textwidth,%
    bbllx=0,bblly=0,bbury=278,bburx=278,clip]
    {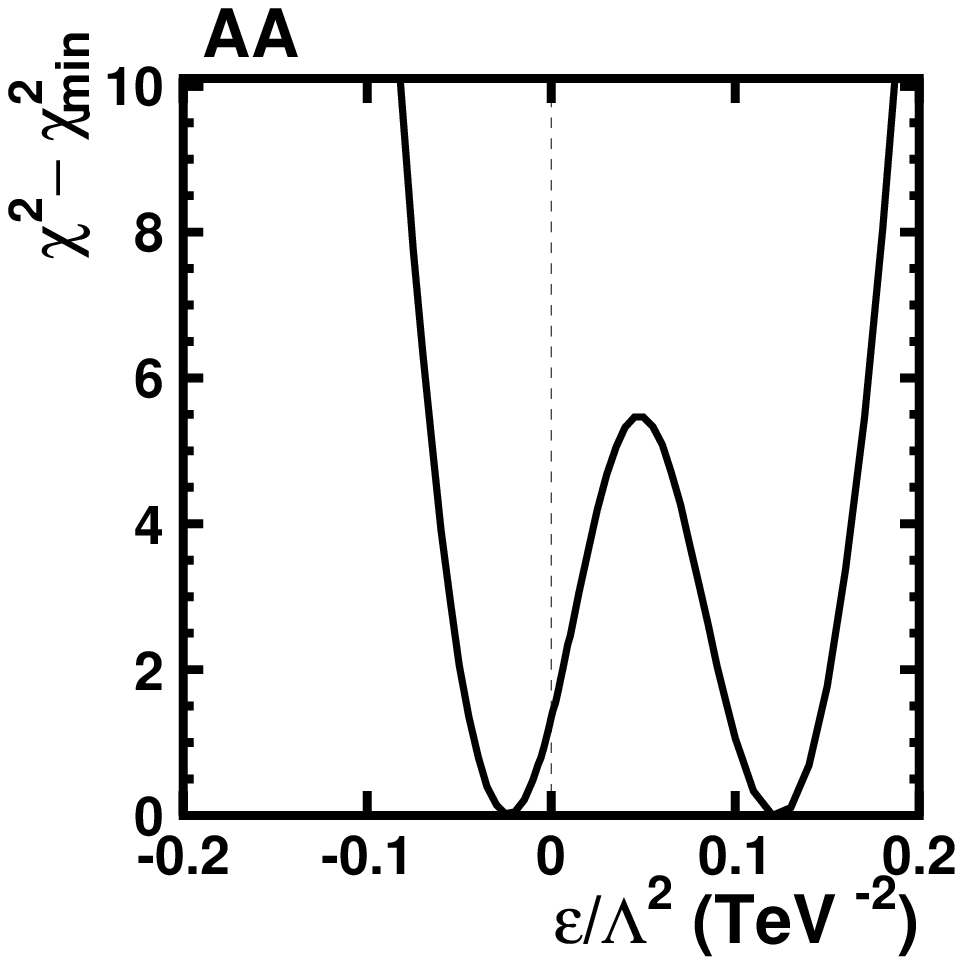}
    \includegraphics[width=0.32\textwidth,%
    bbllx=0,bblly=0,bbury=278,bburx=278,clip]
    {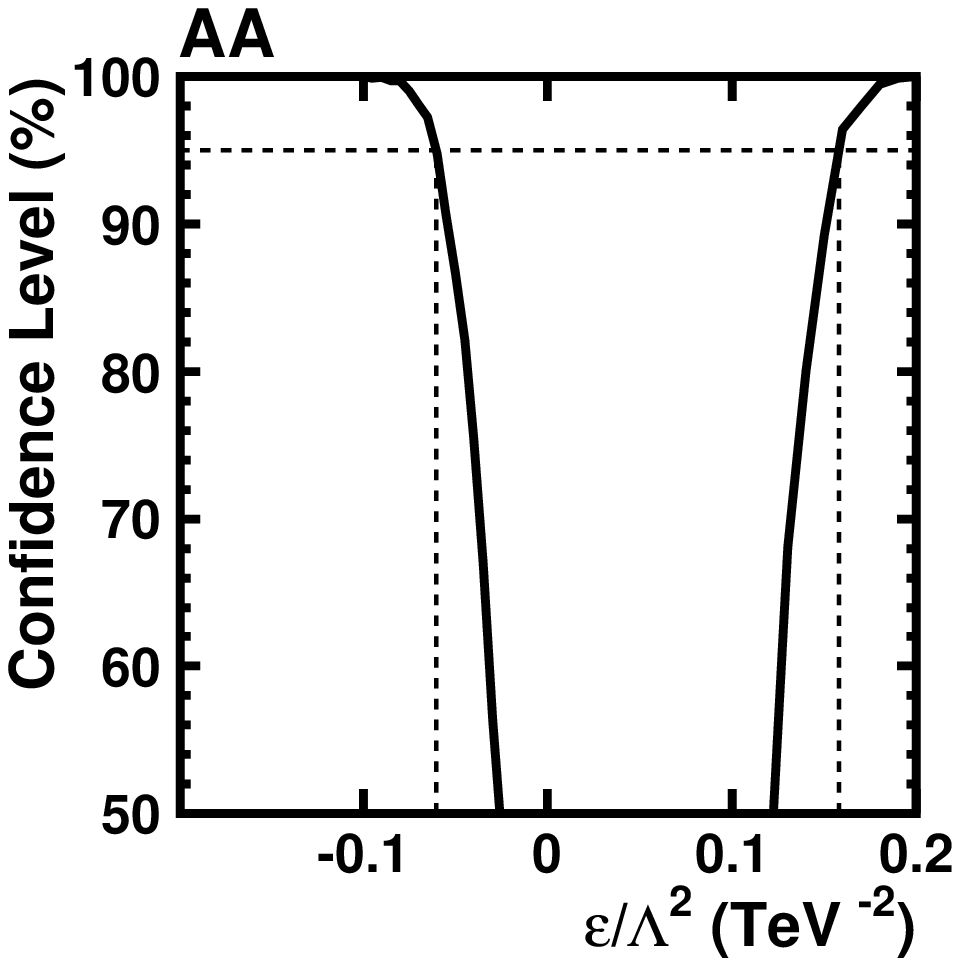}
    \\[5ex]
    \includegraphics[width=0.32\textwidth,%
    bbllx=0,bblly=0,bbury=278,bburx=278,clip]
    {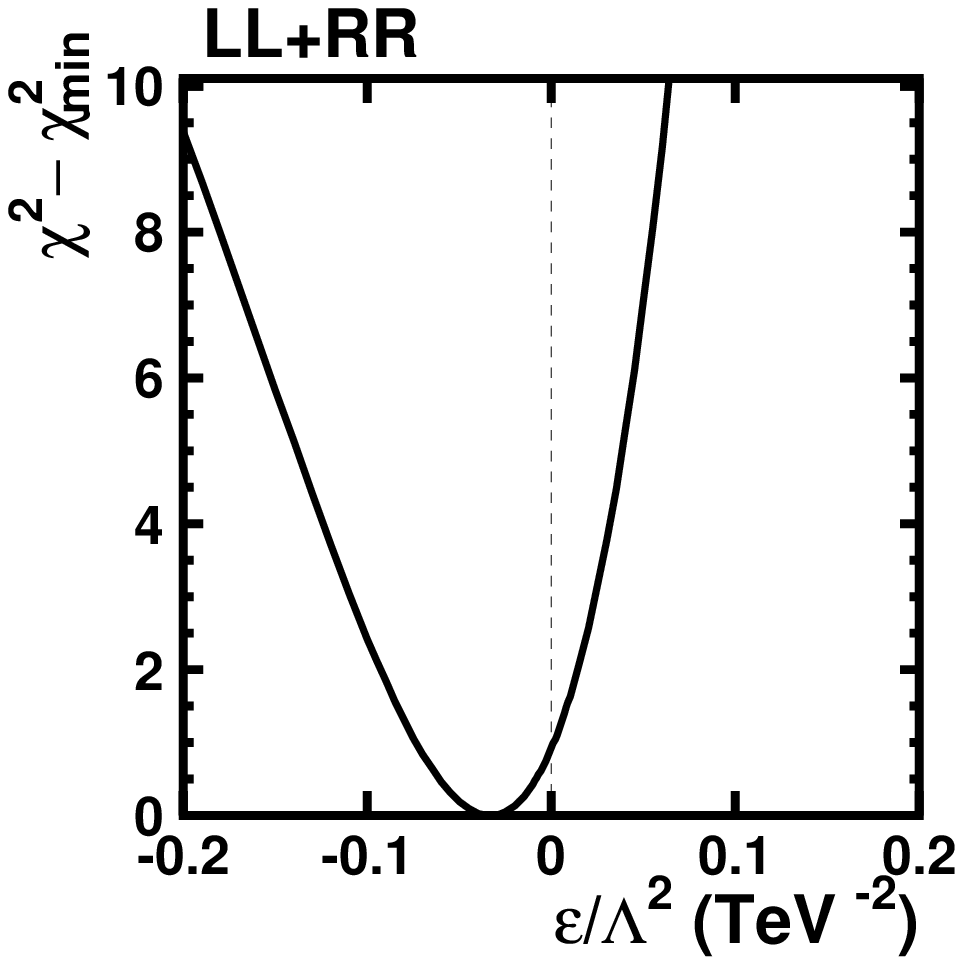}
    \includegraphics[width=0.32\textwidth,%
    bbllx=0,bblly=0,bbury=278,bburx=278,clip]
    {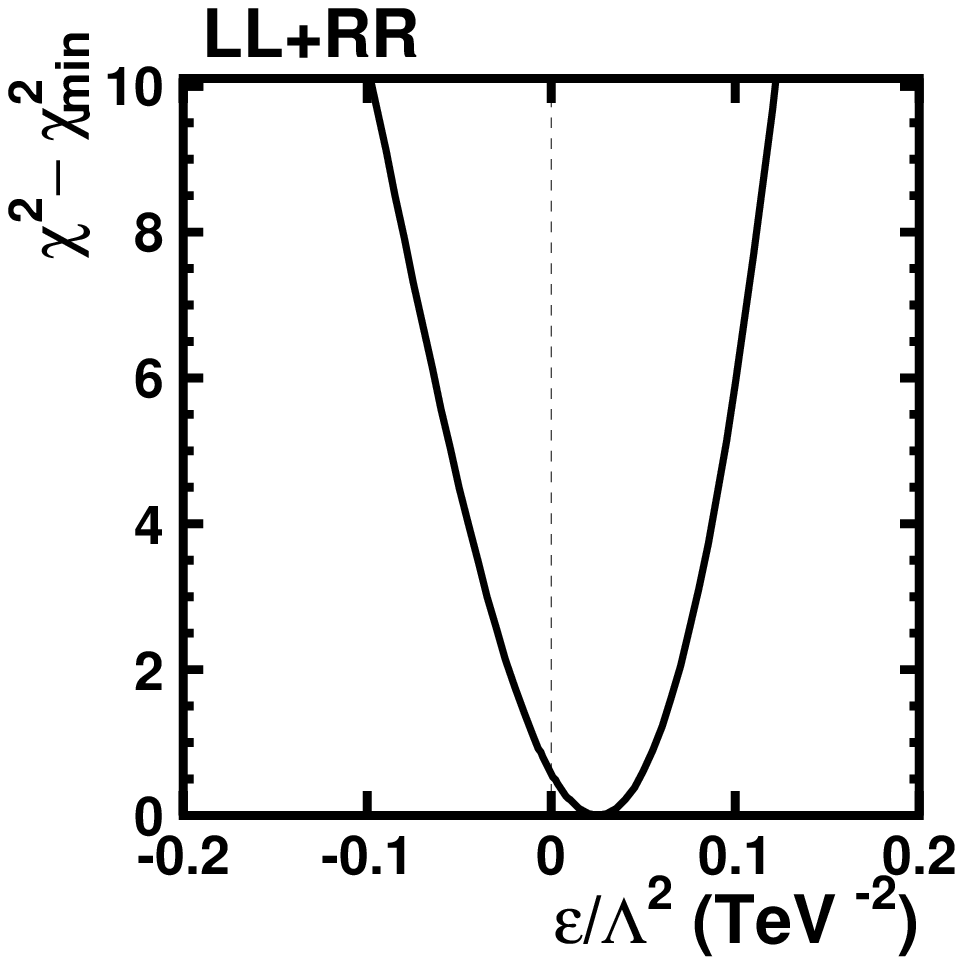}
    \includegraphics[width=0.32\textwidth,%
    bbllx=0,bblly=0,bbury=278,bburx=278,clip]
    {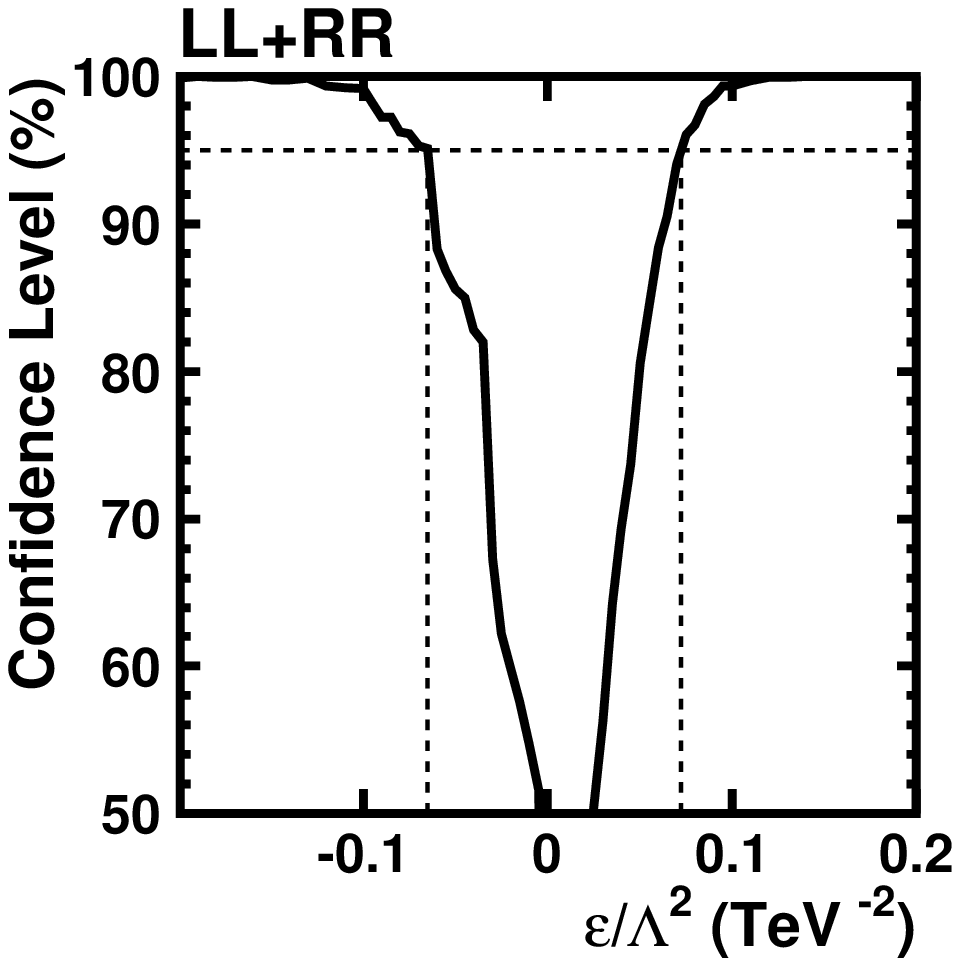}
  \end{center}    
  \caption{Examples of different methods
    to calculate limits on the compositeness scale $\Lambda$ 
    for $VA$, $AA$ and $LL+RR$ models obtained from combined fits
    including all data.
    Distributions of $\chi^2-\chi^2_{min}$ versus $\epsilon/\Lambda^2$
    using statistical errors from the experiment (left)
    and statistical errors from the prediction (center).
    Confidence level versus $\epsilon/\Lambda^2$ from the frequentist
    method with dashed lines indicating the 95\% CL limits for
    positive and negative interference (right).
    In all cases systematics are included.}
  \label{clcurves}
\end{figure}

Taking the statistical errors of the experiment, one can test the 
compatibility of the data with a certain model hypothesis and determine the 
corresponding parameter, for example $1/\Lambda^2$.
Each data point contributes with a fixed weight $\Delta\hat\sigma_i^{-2}$
to the $\chi^2$ function, independent of the model parameter.
However, downward fluctuations with respect to the model predictions
get a larger weight (smaller statistical error) than upward fluctuations
(larger statistical error). 
This property may enforce asymmetric situations and errors, as is 
observed in the present data, see figure~\ref{cieta}.
If the fitted parameter is found to be not significantly different from zero, 
the result can be converted into limits at a given confidence level (CL).
A convenient way is to take the values corresponding to a change of 
$\chi^2(1/\Lambda^{\pm \ 2}) - \chi^2_{SM} = \Delta\chi^2_{CL}$,
for instance $\Delta\chi^2_{95\%} = 3.84$ for a 95\%~CL limit.
This method was used in the previous publication~\cite{h1ci}
and is illustrated for several compositeness models
by the $\chi^2$ distributions in figure~\ref{clcurves}.
This simple and robust definition is certainly meaningful for parabolic curves,
but the probabilistic interpretation becomes ambiguous if there are
secondary minima or other structures. 

An alternative possibility is to assume the validity of a certain CI model 
and to calculate the probability to observe the measured value for a
given model parameter.
Here the statistical error entering the $\chi^2$ function
is taken from the model prediction. This approach is completely
equivalent to using the log-likelihood function, 
and it has been verified that both methods lead to the same results.
Each data point gets a varying weight depending on the model parameter, but
independent of fluctuations in the data. For the present data this 
leads to more symmetric situations and an unbiased evaluation.
In general, the resulting $\chi^2$ curves show a stronger sensitivity to 
details of the CI model, e.g. interference patterns, 
as can be seen in figure~\ref{clcurves}. 
The widths of the distributions are often, but not always, wider than in the 
previous case.
The problem to extract limits from these distributions persists.

The final limits presented here have been determined by applying a 
frequentist approach. 
Starting from a specific model with a scale parameter $\Lambda_{true}$ 
the cross section ${\rm d}\sigma / {\rm d}Q^2$ is calculated
and then smeared according to the statistical error given by the predicted 
number of events.
Distortions due to all uncorrelated and correlated systematic uncertainties 
(except for the parton distributions)
are included, assuming Gaussian behaviour of the errors. 
The Monte Carlo experiment is then analysed in the same way as the 
data, i.e. with statistical errors from the prediction 
and including all systematics,
resulting in a fitted value $\Lambda_{fit}$.
This procedure is repeated numerous times and the fit results are
recorded in a probability distribution.
The scale parameter is then varied and the 95\% confidence lower limit 
$\Lambda^+$ ($\Lambda^-$) is defined as that value $\Lambda_{true}$ 
where 95\% of the Monte Carlo experiments produce  values of $\Lambda_{fit}$ 
which are larger (smaller) than the parameter $\Lambda$ actually obtained 
from the data.
Examples of CL distributions are shown in figure~\ref{clcurves}. 
For the present analysis the frequentist approach provides in general 
slightly weaker and more symmetric limits
than the statistical method used previously~\cite{h1ci}.

%
%

\clearpage

\begin{table}[htb]
\caption{
  Lower limits (95\% CL) on compositeness scale parameters 
  $\Lambda^\pm$ for various chiral structures. 
  Results are given for the present analysis of $e^-p$ and $e^+p$ data 
  and for a combined analysis including $e^+p$ data 
  at $\sqrt{s}=301~\GeV$~\cite{h1xsec}.}
\label{etafits}
\begin{center}
\begin{tabular}{l c c c c c c}
   \hdick \\[-1.5ex]
          & \multicolumn{2}{c}{$e^-p$ $(319\,\GeV)$}  
          & \multicolumn{2}{c}{$e^+p$ $(319\,\GeV)$}  
          & \multicolumn{2}{c}{all $e^+ p$ \& $e^- p$ } \\ 
 coupling & \ $\Lambda^+~[\TeV]$  &  $\Lambda^-~[\TeV]$ \ 
          & \ $\Lambda^+~[\TeV]$  &  $\Lambda^-~[\TeV]$ \ \
          & \ $\Lambda^+~[\TeV]$  &  $\Lambda^-~[\TeV]$ \ \\[1ex]
   \hdick \\[-1.5ex]
 $LL$    & 2.4 & 1.3 & 2.3 & 1.3 &   2.8  & 1.6  \\[.2em]  
 $RR$    & 2.4 & 1.4 & 2.4 & 1.3 &   2.8  & 2.2  \\[.2em] 
 $LR$    & 1.4 & 1.2 & 3.1 & 1.8 &   3.3  & 1.9  \\[.2em]
 $RL$    & 1.4 & 1.2 & 3.1 & 1.9 &   3.3  & 2.0  \\[0.6ex]
 $VV$    & 3.3 & 3.8 & 4.8 & 5.3 &   5.3  & 5.5  \\[.2em]
 $AA$    & 3.2 & 1.6 & 2.3 & 3.7 &   2.5  & 4.1  \\[.2em]
 $VA$    & 2.4 & 2.3 & 2.8 & 2.9 &   2.9  & 3.0  \\[0.6ex]
 $LL+RR$ & 3.1 & 3.4 & 3.1 & 2.3 &   3.7  & 3.9  \\[.2em] 
 $RL+LR$ & 1.8 & 1.4 & 4.2 & 4.0 &   4.4  & 4.4  \\[.2em]
 \hline
\end{tabular}
\end{center}
\end{table}

\clearpage
\begin{table}[htb]
\caption{Coupling coefficients $\eta^q_{ab}$, fermion number $F$ 
  and 95\%~CL lower limits on $M_{LQ}/\lambda$ for scalar ($S$) and vector ($V$) 
  leptoquarks.
  Results are given 
  for the present analysis of $e^-p$ and $e^+p$ data
  and for a combined analysis including $e^+p$ data
  at $\sqrt{s} = 301~\GeV$~\cite{h1xsec}.  
  $L$ and $R$ denote the lepton chirality.
  The subscript $I = 0,\ 1/2,\ 1$ is the weak isospin.
  $\tilde{S}$ and $\tilde{V}$ differ by two units of hypercharge from $S$ and $V$
  respectively.
  Quantum numbers and helicities refer to $e^-q$ and $e^-\bar{q}$ states.}
\label{lqfits}
\begin{center}
\begin{tabular}{l c c c c c c c}
  \hdick \\[-1.5ex]
    & \multicolumn{2}{c}{$\eta^q_{ab}=\epsilon^q_{ab}\cdot (\lambda/M_{LQ})^2$ }  
    &  & $e^-p$ $(319~\GeV)$ & $e^+p$ $(319~\GeV)$   
    & all $e^+ p$ \& $e^-p$ \\[.5ex]
     LQ & $\epsilon^u_{ab}$ & $\epsilon^d_{ab}$ 
    & $F$ & $M_{LQ}/\lambda ~ [ \GeV ]$ 
          & $M_{LQ}/\lambda ~ [ \GeV ]$  
          & $M_{LQ}/\lambda ~ [ \GeV ]$ \\[1ex]
  \hdick \\[-1.5ex]
   $S_0^L$ & 
    \ $\epsilon^u_{LL} = +\frac{1}{2}$ \ & & 2 & 530 & 610 & 710 \\[.2em]
   $S_0^R$ & 
    \ $\epsilon^u_{RR} = +\frac{1}{2}$ \ & & 2 & 480 & 560 & 640 \\[.2em]
   $\tilde{S}_0^R$ & &
    \ $\epsilon^d_{RR} = +\frac{1}{2}$ \   & 2 & 220 & 220 & 330 \\[.2em]
   $S_{1/2}^L$ &
    \ $\epsilon^u_{LR} = -\frac{1}{2}$ \ & & 0 & 290 & 760 & 850 \\[.2em]
   $S_{1/2}^R$ &
    \ $\epsilon^u_{RL} = -\frac{1}{2}$ \   &
    \ $\epsilon^d_{RL} = -\frac{1}{2}$ \   & 0 & 220 & 370 & 370 \\[.2em]
   $\tilde{S}_{1/2}^L$ & &
    \ $\epsilon^d_{LR} = -\frac{1}{2}$ \   & 0 & 200 & 410 & 430 \\[.2em]
   $S_1^L$ &
    \ $\epsilon^u_{LL} = +\frac{1}{2}$ \   &
    \ $\epsilon^d_{LL} = +1$ \             & 2 & 510 & 410 & 490 \\[1ex]
  \hline \\[-1.5ex]
   $V_0^L$ & &
    \ $\epsilon^d_{LL} = -1$ \   & 0 & 480 & 660 & 730 \\[.2em]
   $V_0^R$ & & 
    \ $\epsilon^d_{RR} = -1$ \   & 0 & 420 & 550 & 580 \\[.2em]
   $\tilde{V}_0^R$ & 
    \ $\epsilon^u_{RR} = -1$ \ & & 0 & 810 & 750 & 990 \\[.2em]
   $V_{1/2}^L$ & &
    \ $\epsilon^d_{LR} = +1$ \   & 2 & 280 & 410 & 420 \\[.2em]
   $V_{1/2}^R$ & 
    \ $\epsilon^u_{RL} = +1$ \ &
    \ $\epsilon^d_{RL} = +1$ \   & 2 & 390 & 890 & 950 \\[.2em]
   $\tilde{V}_{1/2}^L$ &
    \ $\epsilon^u_{LR} = +1$ \ & & 2 & 400 & 970 & 1020 \\[.2em]
   $V_1^L$ &
    \ $\epsilon^u_{LL} = -2$ \ &
    \ $\epsilon^d_{LL} = -1$ \   & 0 & 1150 & 960 & 1360 \\[1ex]
  \hline
\end{tabular}
\end{center}
\end{table}

\vspace{1cm}

\begin{table}[htb]
\caption{Coefficients $\epsilon^q_{ab}$
  and 95\%~CL lower limits on $M_{\tilde q}/\lambda'$ for 
  $R_p$ violating couplings to squarks.
  Results are given for the present analysis of $e^-p$ and $e^+p$ data
  and for a combined analysis including $e^+p$ data 
  at $\sqrt{s} = 301~\GeV$~\cite{h1xsec}.}
\label{sqfits}
\begin{center}
\begin{tabular}{l c c c c}
  \hdick \\[-1.5ex]
    &         & $e^-p$ $(319~\GeV)$ & $e^+p$ $(319~\GeV)$   
    & all $e^+ p$ \& $e^-p$ \\[.5ex]
   $\rpv$ \ coupling & $\epsilon^q_{ab}$ 
    & $M_{\tilde q}/\lambda' ~ [ \GeV ]$ 
    & $M_{\tilde q}/\lambda' ~ [ \GeV ]$  
    & $M_{\tilde q}/\lambda' ~ [ \GeV ]$\\[1ex]
  \hdick \\[-1.5ex]
   $\lambda'_{11k}$ \ \ \ $e^+ \bar u \to \bar{\tilde d}^{\,(k)}$ &
    \ $\epsilon^u_{LL} = +\frac{1}{2}$ \ & 530 & 610 & 710 \\[.2em]
   $\lambda'_{1j1}$ \ \ \ $e^+ d \to \tilde{u}^{\,(j)}$ &
    \ $\epsilon^d_{LR} = -\frac{1}{2}$ \   & 200 & 410 & 430 \\[1ex]
  \hline
\end{tabular}
\end{center}
\end{table}

\begin{table}[ht]
\caption{
     Lower limits (95\%~CL) on the gravitational scale $M_S$ assuming
     positive ($\lambda = +1$) and negative ($\lambda = -1$) coupling from
     the present analysis of $e^-p$ and $e^+p$ data 
     and from a combined analysis including $e^+p$ data 
     at $\sqrt{s} = 301~\GeV$~\cite{h1xsec}.}
\label{ledfits}
\begin{center}
\begin{tabular}{l c c c}
   \hdick \\[-1.5ex]
          & \ $e^- p$ $(319~\GeV)$ \ & \ $e^+ p$ $(319~\GeV)$ \
          & \ all $e^+ p$  \& $e^- p$   \\
 coupling $\lambda$ \ & \ $M_S~[\TeV]$ \ & \ $M_S~[\TeV]$ \  & \ $M_S~[\TeV]$ \ 
  \\[1ex]
  \hline \\[-1.5ex]
 $+1$ & 0.58 & 0.77 & 0.82 \\[.2em]
 $-1$ & 0.61 & 0.73 & 0.78 \\[.2em]
 \hline
\end{tabular}
\end{center}
\end{table}

\vspace{1cm}

\begin{table}[ht]
\caption{
     Upper limits (95\%~CL) on the quark radius $R_q$
     assuming point-like leptons ($f_e \equiv 1$) 
     or common form factors ($f_e = f_q$)
     for the present analysis of $e^-p$ and $e^+p$ data 
     and from a combined analysis including $e^+p$ data 
     at $\sqrt{s} = 301~\GeV$~\cite{h1xsec}.}
\label{rqfits}
\begin{center}
\begin{tabular}{l c c c}
   \hdick \\[-1.5ex]
          & \ $e^- p$ $(319~\GeV)$ \ & \ $e^+ p$ $(319~\GeV)$ \
          & \ all $e^+ p$  \& $e^- p$   \\
 form factor \ \ & \ $R_q~[10^{-18}~\m]$ \ 
                 & \ $R_q~[10^{-18}~\m]$ \  & \ $R_q~[10^{-18}~\m]$ \
 \\[1ex]
  \hline \\[-1.5ex]
 $f_e \equiv 1$ & 1.1 & 1.1 & 1.0 \\[.2em]
 $f_e = f_q$    & 0.8 & 0.8 & 0.7 \\[.2em]
 \hline
\end{tabular}
\end{center}
\end{table}

\vfill
\clearpage

%

%

%

%


\begin{thebibliography}{99}

\bibitem{h1ci} C.~Adloff {\it et al.}  [H1 Collaboration],
        Phys.\ Lett.\ B {\bf 479} (2000) 358 [hep-ex/0003002].
\bibitem{h1xsec} C.~Adloff {\it et al.}  [H1 Collaboration],
        Eur.\ Phys.\ J.\ C {\bf 13} (2000) 609 [hep-ex/9908059].
\bibitem{h1xe-p} C.~Adloff {\it et al.}  [H1 Collaboration],
        Eur.\ Phys.\ J.\ C {\bf 19} (2001) 269 [hep-ex/0012052].
\bibitem{h1xe+p} C.~Adloff {\it et al.}  [H1 Collaboration],
         DESY~03--038, 
         submitted to Eur.\ Phys.\ J.\ C,  hep-ex/0304003.

\bibitem{zeusci} J.~Breitweg {\it et al.}  [ZEUS Collaboration],
        Eur.\ Phys.\ J.\ C {\bf 14} (2000) 239 [hep-ex/9905039].

\bibitem{lepci}  R.~Barate {\em et al.} [ALEPH Collaboration],
                Eur. Phys. J. C {\bf 12} (2000) 183 [hep-ex/9904011]; \\
%
                P.~Abreu {\it et al.}  [DELPHI Collaboration],
                Eur.\ Phys.\ J.\ C {\bf 11} (1999) 383;
                Phys.\ Lett.\ B {\bf 485} (2000) 45 [hep-ex/0103025]; \\
%
                M.~Acciarri {\em et al.} [L3 Collaboration], 
                Phys. Lett. B {\bf 489} (2000) 81 [hep-ex/0005028];
                Phys. Lett. B {\bf 470} (1999) 281 [hep-ex/9910056]; \\
%
                G.~Abbiendi {\em et al.} [OPAL Collaboration],
                Eur. Phys. J. C {\bf 13} (2000) 553 [hep-ex/9908008].

\bibitem{tevatronci} F.~Abe {\it et al.}  [CDF Collaboration],
        Phys.\ Rev.\ Lett.\  {\bf 79} (1997) 2192; \\
        B.~Abbott {\it et al.}  [D0 Collaboration],
        Phys.\ Rev.\ Lett.\  {\bf 82} (1999) 4769 [hep-ex/9812010].

\bibitem{cteq} H.-L.~Lai {\it et al.}, 
        Eur.\ Phys.\ J.\ C {\bf 12} (2000) 375  [hep-ph/0201195].

\bibitem{cteq6} J.~Pumplin {\it et al.}, 
       JHEP {\bf 0207} (2002) 012  [hep-ph/0201195].

\bibitem{mrst} A.~D.~Martin, 
        R.~G.~Roberts, W.~J.~Stirling and R.~S.~Thorne,
        Eur.\ Phys.\ J.\ C {\bf 14} (2000) 133  [hep-ph/9907231].

\bibitem{grv}  M.~Gl\"uck, E.~Reya and A.~Vogt, 
        Z.\ Phys.\ C {\bf 67} (1995) 433.


\bibitem{elpr} E.~Eichten, K.~D.~Lane and M.~E.~Peskin,
        Phys.\ Rev.\ Lett.\  {\bf 50} (1983) 811; \\
        R.~R\"uckl,
        Phys.\ Lett.\ B {\bf 129} (1983) 363;
        Nucl.\ Phys.\ B {\bf 234} (1984) 91.

\bibitem{haberl} P.~Haberl, F.~Schrempp and H.-U.~Martyn,
        Proc. Workshop {\em `Physics at HERA'}, 
        eds. W.~Buchm\"uller and G.~Ingelman,
        DESY, Hamburg (1991), vol. 2, p. 1133.


\bibitem{brw} W.~Buchm\"uller, R.~R\"uckl and D.~Wyler,
        Phys.\ Lett.\ B {\bf 191} (1987) 422 
        [Erratum-ibid.\ B~{\bf 448} (1999) 320].


\bibitem{kalinowski} J.~Kalinowski, 
        R.~R\"uckl, H.~Spiesberger and P.~M.~Zerwas,
        Z.\ Phys.\ C {\bf 74} (1997) 595  [hep-ph/9703288].

\bibitem{rpv} J.~Butterworth and H.~Dreiner,
        Nucl.\ Phys.\ B {\bf 397} (1993) 3 [hep-ph/9211204].

\bibitem{add} N.~Arkani-Hamed, S.~Dimopoulos and G.~R.~Dvali,
        Phys.\ Lett.\ B {\bf 429} (1998) 263 [hep-ph/9803315];
       Phys.\ Rev.\ D {\bf 59} (1999) 08600 [hep-ph/9807344].

\bibitem{giudice} G.~F.~Giudice, R.~Rattazzi and J.~D.~Wells,
        Nucl.\ Phys.\ B {\bf 544} (1999) 3
        [corrections in hep-ph/9811291 v2].

\bibitem{cheung} K.~Cheung and G.~Landsberg, 
        Phys.\ Rev.\ D {\bf 65} (2002) 076003 [hep-ph/0110346].

\bibitem{koepp} G.~K\"opp, 
        D.~Schaile, M.~Spira and P.~M.~Zerwas,
        Z.\ Phys.\ C {\bf 65} (1995) 545 [hep-ph/9409457].

\bibitem{pdg} D.~E.~Groom {\it et al.} [Particle Data Group], 
        Eur.\ Phys.\ J.\ C {\bf 15} (2000) 1.

\end{thebibliography}
\end{document}